\documentclass[12pt]{article}
\usepackage{amssymb}

\newcommand{\ppb}{\mbox{\boldmath $P$}}
\newcommand{\xb}{\mbox{\boldmath $x$}}
\newcommand{\pb}{\mbox{\boldmath $p$}}
\newcommand{\hl}{\hat{L}}
\newcommand{\hp}{\hat{P}}
\newcommand{\tpp}{\tilde{P}}
\newcommand{\txx}{\tilde{X}}
\newcommand{\tbeta}{\tilde{\beta}}
\newcommand{\tomega}{\tilde{\omega}}
\newcommand{\tp}{\tilde{p}}

\title{
Lorentz-covariant deformed algebra with minimal length}
\author{C Quesne$^{a}$, V M Tkachuk$^b$\\ 
{\small $^a$ Physique Nucl\'eaire Th\'eorique et Physique Math\'ematique,  Universit\'e
Libre de Bruxelles,} \\  
{\small Campus de la Plaine CP229, Boulevard~du Triomphe, B-1050 Brussels, Belgium}\\
{\small $^b$ Ivan Franko Lviv National University, Chair of Theoretical Physics,}\\
{\small 12, Drahomanov Street, Lviv UA-79005, Ukraine}\\
{\small E-mail: cquesne@ulb.ac.be and tkachuk@ktf.franko.lviv.ua}}
\date{ }
\begin{document}
\baselineskip=20pt plus 1pt minus 1pt
\maketitle

\begin{abstract}
The $D$-dimensional two-parameter deformed algebra with minimal
length introduced by Kempf is generalized to a Lorentz-covariant algebra describing a
($D+1$)-dimensional quantized space-time. For $D=3$, it includes Snyder algebra as a special
case. The deformed Poincar\'e transformations leaving the algebra invariant are identified.
Uncertainty relations are studied. In the case of $D=1$ and one nonvanishing parameter, the
bound-state energy spectrum and wavefunctions of the Dirac oscillator are exactly obtained.
\end{abstract}
\bigskip
%
%
\section{Introduction}

Studies in string theory and quantum gravity suggest the existence of a minimal observable
length. Quantum mechanically this is described by a nonzero minimal uncertainty in position.
Perturbative string theory \cite{gross} actually leads to a generalized uncertainty principle
(GUP) $\Delta X \ge \frac{\hbar}{2} \left(\frac{1}{\Delta P} + \beta \Delta P\right)$, where
$\beta$ is some very small positive parameter. As a consequence $\Delta X \ge \Delta X_0 =
\hbar \sqrt{\beta}$. Such a GUP is also in line with the proposed UV/IR mixing.\par
%
%
This generalized uncertainty relation implies some modification of the canonical
commutation relations (CCR), for which there have been several suggestions. In this
communication, we try to reconcile an old proposal of Snyder, dating back to 1947
\cite{snyder}, with a more recent one, due to Kempf \cite{kempf}. In the former, the
assumption of continuous space-time is abandoned, this leading to a Lorentz-covariant
quantized space-time, where the existence of a natural unit of length is connected with the
non-commutativity of coordinates. In the latter, small quadratic corrections to the CCR are
shown to lead to a GUP of the required form, but the resulting algebra is not Lorentz
covariant. We plan to show here that it is possible to
construct a Lorentz-covariant algebra, reminiscent of that of Kempf and including Snyder
algebra as a special case (for more details see \cite{cq06}).\par
%
%
\section{Deformed algebra generalizing that of Kempf}

In $D$ dimensions, Kempf considered a deformed algebra, whose commutation relations are
given by
\begin{eqnarray}
  && [X^i, P^j] = - {\rm i} \hbar [(1 + \beta \ppb^2) g^{ij} - \beta' P^i P^j], \nonumber \\
  && [X^i, X^j] = {\rm i} \hbar \frac{2\beta - \beta' + (2\beta + \beta') \beta \ppb^2}{1 + 
        \beta \ppb^2} (P^i X^j - P^j X^i), \nonumber \\
  && [P^i, P^j] = 0,  \label{eq:Kempf-com}
\end{eqnarray}
where $X^i$, $P^i$, $i = 1$, 2, \ldots,~$D$, denote the position and momentum coordinates,
while $\beta$ and $\beta'$ are two very small nonnegative deforming parameters
\cite{kempf}. This algebra gives rise to (isotropic) nonzero minimal uncertainties in the
position coordinates $(\Delta X^i)_0 = (\Delta X)_0 = \hbar \sqrt{D\beta + \beta'}$, which
for $D=1$ and $\beta'=0$ reduce to the value deriving from the GUP mentioned in Sec.~1. In
the momentum representation, the deformed position and momentum  operators $X^i$, $P^i$
are represented by
\begin{equation}
  X^i = (1 + \beta \pb^2) x^i + \beta' p^i (\pb \cdot \xb) + {\rm i} \hbar \gamma p^i, 
  \qquad P^i = p^i, \label{eq:Kempf-op}
\end{equation}
where $x^i = {\rm i} \hbar \partial/\partial p^i$ and $\gamma$ is an arbitrary real
constant, which does not enter the commutation relations (\ref{eq:Kempf-com}), but
affects the definition of the scalar product in momentum space.\par
%
%
The algebra (\ref{eq:Kempf-com}) can be converted into a Lorentz-covariant one, by
replacing $\pb^2$ and $\pb \cdot \xb$ in (\ref{eq:Kempf-op}) by the Lorentz-invariant
expressions $\pb^2 - (p^0)^2 = - p_{\nu} p^{\nu}$ and $\pb \cdot \xb - p^0 x^0 = -
p_{\nu} x^{\nu}$, where greek indices run over 0, 1, 2, \ldots,~$D$ and $x^{\nu}$,
$p^{\nu}$ are contravariant vectors in a ($D+1$)-dimensional space-time ($[x^{\mu},
p^{\nu}] = - {\rm i} \hbar g^{\mu\nu}$). Instead of (\ref{eq:Kempf-op}), we therefore
consider the operators
\begin{equation}
  X^{\mu} = (1 - \beta p_{\nu} p^{\nu}) x^{\mu} - \beta' p^{\mu} p_{\nu} x^{\nu} + {\rm
  i} \hbar \gamma p^{\mu}, \qquad P^{\mu} = p^{\mu}, \label{eq:alg-gen}
\end{equation}
which are ($D+1$)-vectors. From (\ref{eq:alg-gen}), we obtain
\begin{eqnarray}
  && [X^{\mu}, P^{\nu}] = - {\rm i} \hbar [(1 - \beta P_{\rho} P^{\rho}) g^{\mu\nu} - \beta'
        P^{\mu} P^{\nu}], \nonumber \\
  && [X^{\mu}, X^{\nu}] = {\rm i} \hbar \frac{2\beta - \beta' - (2\beta + \beta') \beta 
        P_{\rho} P^{\rho}}{1 - \beta P_{\rho} P^{\rho}} (P^{\mu} X^{\nu} - P^{\nu} X^{\mu}),
        \nonumber \\
  && [P^{\mu}, P^{\nu}] = 0,  \label{eq:alg-com}
\end{eqnarray}
where we still assume $\beta$ and $\beta'$ to be two very small nonnegative
parameters.\par   
%
%
The algebra defined in (\ref{eq:alg-com}) can be shown to be left invariant under standard
Lorentz transformations. In the special case where $D=3$ and $\beta = \gamma = 0$, it
reduces to Snyder algebra with $\beta'$ related to the unit of length contained in the latter.
Furthermore, since Kempf algebra, although very similar to (\ref{eq:alg-com}), cannot be
obtained by taking the nonrelativistic limit of the latter, one may say that the algebra
proposed here is a truly new one.\par
%
%
The operators $X^{\mu}$ and $P^{\mu}$ are Hermitian with respect to a new scalar product
in momentum space (in agreement with that of Snyder for $D=3$ and $\beta = \gamma = 0$)
\begin{equation}
  \langle\psi | \phi\rangle = \int \frac{d^D\pb}{[1 - (\beta + \beta') p_{\nu} 
  p^{\nu}]^{\alpha}}\, \psi^*(p^{\mu}) \phi(p^{\mu}),  \label{eq:sc}
\end{equation}
where $\alpha = [2(\beta + \beta')]^{-1} [2\beta + \beta'(D+2) - 2\gamma]$. The weight
function in (\ref{eq:sc}) is free from singularities if physically acceptable states satisfy the
condition 
\begin{equation}
  (\beta + \beta') (p^0)^2 < 1.  \label{eq:cond-accept}
\end{equation}
This means that the energy $E = cP^0 = cp^0$ is not allowed to take very large values
violating condition (\ref{eq:cond-accept}).\par
%
%
\section{Deformed Poincar\'e transformations}

As we already mentioned in Sec.~2, the standard infinitesimal proper Lorentz
transformations $X^{\prime\mu} = X^{\mu} + \delta X^{\mu}$, $P^{\prime\mu} = P^{\mu}
+ \delta P^{\mu}$, where $\delta X^{\mu} = \delta \omega^{\mu}_{\hphantom{\mu}\nu}
X^{\nu}$, $\delta P^{\mu} = \delta \omega^{\mu}_{\hphantom{\mu}\nu} P^{\nu}$ with
$\delta \omega_{\mu\nu} = - \delta \omega_{\nu\mu} \in \mathbf{R}$, leave the algebra
(\ref{eq:alg-com}) invariant. The corresponding generators $\hl_{\alpha\beta} = (1 - 
\beta P_{\rho} P^{\rho})^{-1} (X_{\alpha} P_{\beta} - X_{\beta} P_{\alpha})$, such that 
$\delta X^{\mu} = [{\rm i}/(2\hbar)] \delta \omega^{\alpha\beta}
[\hl_{\alpha\beta}, X^{\mu}]$ and $\delta P^{\mu} = [{\rm i}/(2\hbar)]
\delta \omega^{\alpha\beta} [\hl_{\alpha\beta}, P^{\mu}]$, are however
deformed, although they satisfy the standard so($D$,1) commutation relations
\[
  \left[\hl_{\alpha\beta}, \hl_{\rho\sigma}\right] = - {\rm i} \hbar \left(g_{\alpha\rho} \hl_{\beta\sigma}
  - g_{\alpha\sigma} \hl_{\beta\rho} - g_{\beta\rho} \hl_{\alpha\sigma} + g_{\beta\sigma}
  \hl_{\alpha\rho}\right).
\]
\par
%
%
The invariance of (\ref{eq:alg-com}) under proper Lorentz transformations can be extended to
improper ones as it can be checked from the action of the discrete symmetries, namely parity
$P$ and time reversal $T$.\par
%
%
Infinitesimal translations also leave (\ref{eq:alg-com}) invariant provided they are deformed as
$X^{\prime\mu} = X^{\mu} + \delta X^{\mu}$, $P^{\prime\mu} = P^{\mu}$,
where $\delta X^{\mu} = - \delta a^{\mu} - g(P_{\rho} P^{\rho}) \delta a_{\nu} P^{\nu}
P^{\mu}$ with $\delta a^{\mu} \in \mathbf{R}$ and $g(P_{\rho} P^{\rho}) = (1 - \beta
P_{\rho} P^{\rho})^{-2} [2\beta - \beta' - (2\beta + \beta') \beta P_{\rho} P^{\rho}]$. With
these new definitions, the deformed generators $\hp_{\alpha} = (1 - \beta P_{\rho}
P^{\rho})^{-1} P_{\alpha}$, such that $\delta X^{\mu}  = (\rm i/\hbar) \delta a^{\alpha}
[\hp_{\alpha}, X^{\mu}]$ and $\delta P^{\mu} = (\rm i/\hbar) \delta a^{\alpha}
[\hp_{\alpha}, P^{\mu}]$, fulfil the standard commutation relations 
\[
  \left[\hp_{\alpha}, \hp_{\beta}\right] = 0, \qquad \left[\hl_{\alpha\beta}, \hp_{\rho}
  \right] = {\rm i} \hbar \left(g_{\beta\rho} \hp_{\alpha} - g_{\alpha\rho}
  \hp_{\beta}\right). 
\]
\par
%
%
We conclude that the deformed operators $\hl_{\alpha\beta}$ and $\hp_{\alpha}$
provide us with a realization of the conventional Poincar\'e algebra iso($D$,1) leaving
(\ref{eq:alg-com}) invariant.\par
%
%
\section{Uncertainty relations for position and momentum}

Since the algebra (\ref{eq:alg-com}) is invariant under rotations, we may choose any pair of
position and momentum components $X^i$, $P^i$ ($i \in \{1, 2, \ldots, D\}$) to determine
the deformed uncertainty relation:
\[
  \Delta X^i \Delta P^i \ge \frac{\hbar}{2} \left|1 - \beta \left\{\langle (P^0)^2 \rangle - 
  \sum_{j=1}^D \left[(\Delta P^j)^2 + \langle P^j \rangle^2\right] \right\} + \beta'
  \left[(\Delta P^i)^2 + \langle P^i \rangle^2\right]\right|.  \label{eq:UR}
\]
\par
%
%
If, for simplicity's sake, we assume isotropic uncertainties $\Delta P^j = \Delta P$, $j=1$,
2,~\ldots, $D$, we get for each $X^i$ an uncertainty relation similar to the GUP in
Sec.~1, from which it results that $\Delta X^i$ has a nonvanishing minimum
\[
  \Delta X^i_{\rm min} = \hbar \sqrt{(D\beta + \beta') \left\{1 - \beta \left[\langle (P^0)^2 
  \rangle - \sum_{j=1}^D \langle P^j \rangle^2\right] + \beta' \langle P^i \rangle^2 \right\}}
\]
provided the quantity between curly brackets on the right-hand side is positive. This condition 
is always satisfied by physically acceptable states due to Eq.~(\ref{eq:cond-accept}). We
therefore arrive at an isotropic absolutely smallest uncertainty in position given by
\[
  (\Delta X)_0 = (\Delta X^i)_0 = \hbar \sqrt{(D\beta + \beta') \left[1 - \beta \langle 
  (P^0)^2 \rangle \right]}. 
\]
As compared with Kempf's result, there is an additional factor $\sqrt{1 - \beta \langle
(P^0)^2 \rangle}$ reducing $(\Delta X)_0$.\par
%
%
\section{\boldmath Application to the ($1+1$)-dimensional Dirac oscillator}

The $3+1$-dimensional Dirac oscillator (DO) was introduced a long time ago \cite{ito}, but its
name was only coined later on \cite{moshinsky}.  This system has aroused a lot of interest
both because it is one of the few examples of exactly solvable Dirac equation and because it
can be applied to a lot of physical problems. As it is the only relativistic problem that has been
solved with Kempf algebra \cite{cq05} (see also \cite{nouicer} for the $1+1$-dimensional
case), it is interesting to see what is the effect of the deformed algebra (\ref{eq:alg-com}) on
this system.\par
%
%
{}For simplicity's sake, we consider here the simplifying assumptions $D=1$ and $\beta' = 0$.
On taking for the Dirac $2 \times 2$ matrices $\hat{\alpha}_x$ and $\hat{\beta}$ the
standard Pauli spin matrices $\sigma_x$ and $\sigma_z$, respectively, and going to
dimensionless operators $\txx^{\mu} = X^{\mu}/a$, $\tpp^{\mu} = (a/\hbar) P^{\mu}$, 
$\mu = 0$, 1, where $a = \hbar/(mc)$, the DO equation reads in momentum
representation as
\begin{equation}
  (\sigma_x \tpp - \tomega \sigma_y \txx + \sigma_z) \psi(\tp, \tp^0) = \tpp^0 \psi(\tp, 
  \tp^0), \label{eq:DO-ter} 
\end{equation}
where 
\begin{eqnarray*}
  && \tpp^0 = \tp^0 \qquad \tpp = \tp, \\
  && \txx^0 = - {\rm i} f(\tp, \tp^0) \frac{\partial}{\partial \tp^0}, \quad \txx = {\rm i}
       f(\tp, \tp^0) \frac{\partial}{\partial \tp}, \quad f(\tp, \tp^0) = 1 - \tbeta [(\tp^0)^2 -
       \tp^2], 
\end{eqnarray*}
and $\tp^{\mu} = (a/\hbar) p^{\mu}$, $\tomega = \hbar\omega/(mc^2)$, $\tbeta =
\beta m^2 c^2$.\par
%
%
On separating the wavefunction $\psi = \left(\begin{array}{c} \psi_1 \\ \psi_2 \end{array}
\right)$ into large $\psi_1$ and small $\psi_2$ components and defining
\[
  B^{\pm} = \tpp \pm {\rm i} \tomega \txx = \tp \mp \tomega f(\tp, \tp^0) 
  \frac{\partial}{\partial \tp},
\]
the DO equation (\ref{eq:DO-ter}) can be written as two coupled equations
\begin{equation}
  B^+ \psi_2(\tp, \tp^0) = (\tp^0 - 1) \psi_1(\tp, \tp^0), \qquad
  B^- \psi_1(\tp, \tp^0) = (\tp^0 + 1) \psi_2(\tp, \tp^0). \label{eq:coupled}
\end{equation}
These lead to two factorized equations
\begin{equation}
  B^+ B^- \psi_1(\tp, \tp^0) = e(\tp^0) \psi_1(\tp, \tp^0), \qquad
  B^- B^+ \psi_2(\tp, \tp^0) = e(\tp^0) \psi_2(\tp, \tp^0), \label{eq:factorized}
\end{equation}
for the large and small components, respectively. Here $e(\tp^0) \equiv (\tp^0)^2 - 1$.\par
%
%
Such factorized equations have the form of energy-eigenvalue equations in
SUSYQM~\cite{cooper}. It should be stressed, however, that the starting Hamiltonian $H =
B^+ B^-$ depends on $(\tp^0)^2$, which determines the eigenvalues $e(\tp^0)$. Hence, in
contrast with conventional SUSYQM, $H$ is energy dependent. We have however shown that
provided one proceeds with care, the shape-invariance method can be used to determine the
possible values of $e(\tp^0)$. The result reads $e_n(\tp^0) = \tomega n (2 +
\tbeta\tomega n) [1 - \tbeta (\tp^0)^2].$\par 
%
%
{}For such an expression to be compatible with the definition of $e(\tp^0)$, given below
(\ref{eq:factorized}), $\tp^0$ must be quantized and its allowed values given by
\[
  \tp^0_{n, \tau} = \tau \left(\frac{1 + \tomega n (2 + \tbeta\tomega n)}{1 + \tbeta
  \tomega n (2 + \tbeta\tomega n)}\right)^{1/2} = \frac{\tau}{\sqrt{\tbeta}} \left(1 +
  \frac{\tbeta - 1}{(1 + \tbeta\tomega n)^2}\right)^{1/2},  \label{eq:p^0} 
\]
where $\tau = \pm 1$ and $n$ may, in principle, run over $n=0$, 1, 2,~\ldots. Since $E = c 
p^0 = m c^2 \tp^0$, the Dirac oscillator energy spectrum is given by
\begin{equation}
  E_{n, \tau} = \frac{\tau c}{\sqrt{\beta}} \left(1 + \frac{\beta m^2 c^2 - 1}{(1 + \beta
  m\hbar\omega n)^2}\right)^{1/2},  \label{eq:E}
\end{equation}
where the values of $(n, \tau)$ depend on the existence of normalizable solutions to 
Eq.~(\ref{eq:coupled}). It can actually be proved that $n=0$, 1, 2,~\ldots, or
$n=1$, 2,~\ldots, according to whether $\tau = +1$ or $\tau = -1$.\par
%
%
Eq.~(\ref{eq:E}) has three important consequences. Firstly, it shows that we have to restrict
ourselves to deformations such that $\beta < 1/(m^2c^2)$, because otherwise we
would get decreasing values of $|E_{n,\tau}|$ when $n$ increases --- an unphysical behaviour
for an energy spectrum. This is also confirmed by the existence of well-behaved
wavefunctions in such a range of $\beta$ values. Secondly, the energy spectrum turns out to
be bounded ($mc^2 \le |E_{n,\tau}| < c/\sqrt{\beta}$) in contrast with what happens 
for the conventional DO. It can actually be shown that in the limit where $\beta$ vanishes,
the usual unbounded energy spectrum is retrieved. Thirdly, in the nonrelativistic limit
$\hbar\omega/(mc^2) \ll 1$, we get the spectrum obtained with Kempf algebra
\cite{nouicer}, except for the presence of the additional factor $(1 + \beta m\hbar\omega
n)^{-1}$ in the energies. This is another evidence of the novelty of the Lorentz-covariant
deformed algebra considered here, as compared with Kempf one.\par
%
%
The wavefunctions associated with the spectrum (\ref{eq:E}) can be found in two
steps. First, SUSYQM techniques are used to determine the eigenfunctions of $H$ and of its
SUSY partner, satisfying Eq.~(\ref{eq:factorized}). From them, the large and small
components, $\psi_1^{(n,\tau)}(\tp) \equiv \psi_1(\tp,\tp^0_{n,\tau})$ and
$\psi_2^{(n,\tau)}(\tp) \equiv \psi_2(\tp,\tp^0_{n,\tau})$, fulfilling
Eq.~(\ref{eq:coupled}), are then obtained. Note that the latter step is not trivial because
though the solutions of (\ref{eq:coupled}) provide us with solutions to (\ref{eq:factorized}),
the converse is not necessarily true. As a result of the calculations, we get wavefunctions
normalized with respect to the scalar product (\ref{eq:sc}), but since the weight function
$1/f(\tp,\tp^0_{n,\tau})$ in (\ref{eq:sc}) is energy dependent, the standard orthogonality
relation between separate bound states is lost. This problem is one of the many known puzzles
inherent in the use of energy-dependent Hamiltonians in quantum mechanics (for a recent
review see \cite{formanek}).\par
%
%
In conclusion, we would like to mention a few remaining problems for future
investigations: (i) a thorough analysis of the time-energy uncertainty, required by its special
status in quantum mechanics; (ii) an attempt to restore some of the properties of ordinary
quantum mechanics that are spoilt by the energy dependence of the Dirac oscillator
Hamiltonian; and (iii) a search for a physical interpretation of the deforming parameter
$\beta$, taking into account that $\beta'$, being related to Snyder natural unit of length,
has already received one \cite{kowalski}.\par          
%
%
\bigskip
C.Q.\ is a Research Director of the National Fund for Scientific Research (FNRS), Belgium.\par
%

\end{document}